\documentstyle[aps,epsfig,prl,twocolumn]{revtex}
\def\be{\begin{equation}}
\def\ee{\end{equation}}
\def\bea{\begin{eqnarray}}
\def\eea{\end{eqnarray}}

\def\c{{\cal C}}

\newcommand{\eins}{\mbox{$1 \hspace{-1.0mm}  {\bf l}$}}
\newcommand{\bra}[1]{\mbox{$\langle #1 |$}}
\newcommand{\ket}[1]{\mbox{$| #1 \rangle$}}
\newcommand{\braket}[2]{\mbox{$\langle #1  | #2 \rangle$}}
\newcommand{\proj}[1]{\ket{#1}\!\bra{#1}}
\newcommand{\ope}[2]{\ket{#1}\!\bra{#2}}

\begin{document}
\draft

\title{Optimal distillation of a GHZ state}

\author{ A. Ac\'{\i}n$^1$, E. Jan\'{e}$^1$, W. D\"ur$^2$ and G. Vidal$^2$}

\address{
$^1$ Departament d'Estructura i Constituents de la Mat\`eria, Universitat de Barcelona, E-08028 Barcelona, Spain.\\
$^2$ Institut f\"ur Theoretische Physik, Universit\"at Innsbruck, A-6020 Innsbruck, Austria.}

\date{\today}

\maketitle

\begin{abstract}
We present the optimal local protocol to distill a  Greenberger-Horne-Zeilinger (GHZ) state from a single copy of any pure state of three qubits. 
\end{abstract}



Because of its relevance in quantum information theory, entanglement has been attracting considerable attention in recent years. The initial efforts, mainly devoted to acquire a quantitative description of bipartite quantum correlations, led to the identification of the measures governing pure-state entanglement both in the asymptotic \cite{Ben96} and non-asymptotic \cite{Vid99} regimes. As a result, it is nowadays known how a bipartite system prepared in a pure state can be optimally manipulated under the restriction that only local operations on the subsystems aided with classical communication (LOCC) are allowed.

Once a relatively complete command on pure-state entanglement had been achieved for bipartite systems, efforts have moved quite recently to address the somehow more intrincate case of tripartite entanglement, for which a system consisting of three two-level subsystems ---that is, a three-qubit system--- is the simplest scenario.
 
There are two distinct ways in which three qubits can be entangled \cite{Dur00}, in the following sense. Let us identify as equivalent all pure states that can be reversibly interconverted, with some finite probability of success, when the parties are allowed to perform LOCC. Then, whereas all entangled pure states of two qubits are equivalent to the Einstein-Podolsky-Rosen (EPR) state \cite{EPR},
$ 1/\sqrt{2}(\ket{00} + \ket{11}),$
it turns out that truly tripartite pure-state entanglement of three qubits is either equivalent to the GHZ state \cite{GHZ},
\be
\ket{GHZ} \equiv \frac{1}{\sqrt{2}}(\ket{000} + \ket{111}),
\label{GHZ}
\ee  
or else to the W state \cite{Dur00},
$1/\sqrt{3}(\ket{001} + \ket{010} + \ket{100}),$
with these two states being completely inaccessible from each other by means of LOCC. 

 Although it is genuinely tripartite, the entanglement of the $W$ state only maximizes two-qubit quantum correlations \cite{Dur00}. As far as three-qubit correlations are concerned, the GHZ state appears as the maximally entangled state \cite{Gis98}: it violates Bell inequalities maximally and it maximizes the mutual information of local measurements. It is also the only state from which an EPR state between any two chosen qubits can be obtained with certainty. 

In this Letter we are concerned with the distillation of a GHZ state from an arbitrary pure state $\ket{\psi}$ of three qubits (see also \cite{Coh00}). We present the optimal strategy to distill a GHZ state starting from a single copy of $\ket{\psi}$. That is, we provide the local protocol that allows the three parties to transform the state $\ket{\psi}$ into a GHZ state with the greatest a priori probability of success compatible with LOCC. Marginally, we also discuss the best approximate transformation \cite{Vid00} of $\ket{\psi}$ into $\ket{GHZ}$ by means of LOCC, which turns out to simply consist in a concatenation of local unitary transformations. Obviously, these results may be of practical interest in any situation where three parties wishing to have a GHZ state only share the state $\ket{\psi}$, and where for some reason ---for instance because they are in sufficiently distant locations--- the restriction to perform only LOCC is not merely academical. From a more general perspective, several difficulties have been faced which stem from characteristic features of entanglement involving more than two subsystems, such as the non-existence of the Schmidt decomposition (that is, from the fact that not all three-qubit states can be brought, by means of local unitary transformations, into the form $\alpha\ket{000} + \beta \ket{111}$, see \cite{des}). In this sense, our work explores the nature of entanglement beyond the bipartite case, while providing useful tools for its study.

Let us first briefly review how, in the two-party setting, Alice and Bob can optimally distill a generic pure state of two qubits into an EPR state. The initial state can be transformed, by means of local unitary operations on Alice and Bob's sides, into a reference state whose Schmidt decomposition is
\be
\alpha\ket{00} + \beta \ket{11},
\label{2q}
\ee
where $\alpha \geq \beta \geq 0$ and $\alpha^2+\beta^2=1$. Then, following \cite{Ben96}, the diagonal operator
$M\equiv \frac{\beta}{\alpha} \proj{0} + \proj{1}$ 
---corresponding to a POVM \cite{rab} on, say, Alice's part--- transforms state (\ref{2q}) into an EPR state with probability $2\beta^2$. Finally, that such distillation protocol is optimal \cite{Lo97} can be interpreted in terms of the entanglement monotone $E_2\equiv \beta^2$ \cite{Vid99}. Indeed, if a protocol would distill the EPR state with greater probability, this would contradict the non-increasing character of $E_2$ under LOCC.

 Similar steps will be followed in order to find which is the optimal distillation protocol of a GHZ state in a three-qubit system: ($i$) the Schmidt decomposition will be replaced with a convenient two-term product decomposition; ($ii$) a subclass of distillation protocols, namely those in which each party is only allowed to perform one positive operator-valued measurement (POVM) and only one of the overall outcomes is a GHZ state, will be discussed and optimized; ($iii$) the optimal of such {\em one succesful branch protocols} (OSBP) will be proved to be the best distillation protocol by showing that the associated probability of success is an entanglement monotone \cite{Vid99b}.  

Recall that a necessary and sufficient condition for the distillability of a GHZ from a three-qubit pure state $\ket{\psi}$ is that the three reduced density matrices $\rho_A\equiv$Tr$_{BC}\proj{\psi}$, $\rho_B$ and $\rho_C$ have rank $2$, and that the range of $\rho_{BC}\equiv$Tr$_{A}\proj{\psi}$ contains two product vectors, say $\ket{b_1c_1}$ and $\ket{b_2c_2}$ \cite{Dur00}. Only in this case $\ket{\psi}$ admits a (unique) two-term product decomposition,
\be
\label{state}
\ket{\psi}=\mu_1\ket{a_1b_1c_1}+\mu_2 e^{i \varphi}\ket{a_2b_2c_2},  ~~~\mu_1 \geq \mu_2 > 0,
\ee
where $\varphi \in [0,2\pi)$ and the normalized local vectors $\ket{\kappa_1},\ket{\kappa_2} \in \c^2$ satisfy $1>\braket{\kappa_1}{\kappa_2}\geq 0$, $\kappa = a,b,c$. In its turn, a distillation protocol consists of a series of local POVMs, where the specific POVM to be performed at any time may be choosen depending on previous outcomes, the whole protocol having a tree structure. For a given branch of outcomes the initial state $\ket{\psi}$ is transformed into a final state 
\be
\label{prot}
A\otimes B\otimes C\,\ket{\psi},
\ee
where the local operator $A$ collects contributions from all POVMs performed by Alice in that branch, and similarly for operators $B$ and $C$. Characterizing operator $A$ by how it transforms $\ket{a_1}$ and $\ket{a_2}$,
\be
A = \alpha_1\ope{a'_1}{\tilde{a}_1}+ \alpha_2 e^{i \varphi_a}\ope{a_2'}{\tilde{a}_2},   
\label{opeA0}
\ee
where $\alpha_1,\alpha_2,\braket{a_1'}{a_2'} $$\geq$$ 0$ and $\{\ket{\tilde{a}_1},\ket{\tilde{a}_2}\}$ is the biorthonormal basis to $\{\ket{a_1},\ket{a_2}\}$, i.e. $\braket{\tilde{a}_i}{a_j}= \delta_{ij}$,
we observe that the modifications introduced by A in $\ket{\psi}$, $
A\otimes\eins_B\otimes\eins_C \ket{\psi} = \alpha_1\mu_1\ket{a'_1b_1c_1}+\alpha_2\mu_2 e^{i (\varphi+\varphi_a)}\ket{a_2'b_2c_2},$
concern the weights $\mu_1$ and $\mu_2$, the relative phase $e^{i\varphi}$ and the local scalar product $\braket{a_1}{a_2}$, whereas the other two local scalar products, $\braket{b_1}{b_2}$ and $\braket{c_1}{c_2}$, remain unchanged.
  We can therefore readily conclude, from the uniqueness of the product decomposition (\ref{state}) and the previous observation, that the only way for the resulting state (\ref{prot}) to be proportional to $\ket{GHZ}$ is that ($i$) each local operator transforms the corresponding couple of non-orthogonal, local states into an orthonormal pair,
$\{\ket{\kappa_1},\ket{\kappa_2}\} \longrightarrow \{\ket{0},\ket{1}\},~\kappa=a,b,c,$
that is,
\bea
A &=& \alpha_1\ope{0}{\tilde{a}_1}+ \alpha_2 e^{i \varphi_a}\ope{1}{\tilde{a}_2},\nonumber\\
B &=& \beta_1\ope{0}{\tilde{b}_1}+ \beta_2 e^{i \varphi_b}\ope{1}{\tilde{b}_2},  \label{opeA}\\
C &=& \gamma_1\ope{0}{\tilde{c}_1}+ \gamma_2 e^{i \varphi_c}\ope{1}{\tilde{c}_2},\nonumber
\eea
and that ($ii$) their combined effect modifies the weights $\mu_1$ and $\mu_2$ to be equal,
$\mu_i \longrightarrow\alpha_i\beta_i\gamma_i\mu_i=\sqrt{p/2},~~i=1,2,\label{mu}$\cite{phase},
where 
\be
p \equiv \bra{\psi}A^{\dagger}A\otimes B^{\dagger}B\otimes C^{\dagger}C\ket{\psi} = 2(\alpha_1\beta_1\gamma_1\mu_1)^2
\label{prob}
\ee
is the probability that the distillation protocol follows that particular, successful branch.

 We have thus seen that the three parties must act on the system in order to distill a GHZ, because each of them must orthonormalize its local couple of states. The other relevant feature of the distillation process, namely making the relative weights equal, may be distributed in several ways among the parties. 

 Our OSBP for distillation consists in each of the parties performing a unique, two-outcome POVM, and it is built in such a way that after each POVM one of the two possible resulting states contains no three-partite entanglement at all, so that only one branch of the whole protocol succeeds in distilling a GHZ (see figure 1). In mathematical terms this implies that if e.g. $A$ in (\ref{opeA}) is the successful operator in Alice's POVM, then the other operator of the local POVM, $\bar{A}$, which satisfies
$A^{\dagger}A + \bar{A}^{\dagger}\bar{A}=\eins_A,$ 
has rank equal to 1, so that the resulting state $\bar{A}\otimes\eins_B\otimes\eins_C \ket{\psi}$ is product in Alice's subsystem. Expressing the identity operator as
\be
\eins_A = \sum_{i,j} \braket{a_i}{a_j} \ope{\tilde{a}_i}{\tilde{a}_j},
\ee
and requiring that det[$\eins_A-A^{\dagger}A]=0$, we find 
\be
(1-\alpha_1^2)(1-\alpha_2^2)=\braket{a_1}{a_2}^2.
\label{constA}
\ee
Operators $B$ and $C$ are also accordingly constrained by
\bea
(1-\beta_1^2)(1-\beta_2^2)=\braket{b_1}{b_2}^2,
\nonumber\\
(1-\gamma_1^2)(1-\gamma_2^2)=\braket{c_1}{c_2}^2.
\label{constC}
\eea
To completely characterize our protocol, we will further impose that it succeeds with the greatest possible probability over all OSBPs. The optimal OSBP probability,
\be
P(\psi)\equiv \max_{\alpha_1,\beta_1,\gamma_1} 2 (\alpha_1\beta_1\gamma_1\mu_1)^2,
\ee
---where the constraints  $\alpha_1\beta_1\gamma_1\mu_1 = \alpha_2\beta_2\gamma_2\mu_2$ and (\ref{constA}-\ref{constC}) hold--- reads
\bea
\label{prob3}
&P&=^{max}_{x>0}\quad \frac{f_1(x)f_2(x)}{2}\left(1-\sqrt{1-\frac{4(1-\braket{a_1}{a_2}^2)}{f_1^2(x)}}\right) \nonumber\\
&\times&\left(1-\sqrt{1-\frac{4\mu_1^2\mu_2^2(1-\braket{b_1}{b_2}^2)(1-\braket{c_1}{c_2}^2)}{f_2^2(x)}}\right) ,
\eea
where
\bea
&&f_1(x)\equiv (x^2+1)/x, \nonumber\\
&&f_2(x)\equiv (\mu_2^2x^2+2\mu_1\mu_2 \braket{b_1}{b_2}\braket{c_1}{c_2}x+\mu_1^2)/x .
\eea
 The maximization in (\ref{prob3}) involves a polynomial equation of degree 6, which in general requires numerical calculations. For illustrative purposes and later reference, we will next consider two particular situations which can be solved analytically. 

Suppose, first, that the parties share the state
\be
\ket{\phi^1}=\mu_1\ket{0 0 c_1}+\mu_2\ket{1 1 c_2},~~~~~\mu_1\geq\mu_2>0
\ee 
that is, with the local states already orthogonal in Alice's and Bob's subsystems. Then the previous optimization over all OSBP leads to a maximal probability
\be
P(\phi^1)=1-\sqrt{1-4\mu_1^2\mu_2^2 (1-\braket{c_1}{c_2}^2)}.
\ee
This probability is two times the smallest eigenvalue of $\rho_C$, which has a decreasing behaviour under LOCC. Monotonicity of this eigenvalue implies therefore that $P(\phi^1)$ is also the maximal probability for any general distillation protocol. It turns out that Claire ---i.e. the party that has to orthonormalize its local states--- is the only one that needs to perform a local POVM.

The second particular case concerns states of the form
\be
\label{cas2}
\ket{\phi^2}=\mu_1\ket{0 b_1 c_1}+\mu_2\ket{1 b_2 c_2},~~~~~\mu_1\geq\mu_2>0
\ee 
that is with orthogonal states in Alice's subsystem. The maximum probability for OSBP is
\bea
&P&(\phi^2)=(1+2\mu_1\mu_2 \braket{b_1c_1}{b_2c_2})\nonumber\\
&\times& \left(1-\sqrt{1-\frac{4\mu_1^2\mu_2^2 (1-\braket{b_1}{b_2}^2)(1-\braket{c_1}{c_2}^2)}{(1+2\mu_1\mu_2\braket{b_1c_1}{b_2c_2})^2}}\right).
\label{prob2}
\eea
Again, only Bob and Claire need to act on the system, with the corresponding POVMs satisfying
\be
\frac{\beta_1^2}{\beta_2^2} = \frac{\mu_1}{\mu_2}\frac{\mu_1\braket{b_1}{b_2}+\mu_2\braket{c_1}{c_2}}{\mu_2\braket{b_1}{b_2}+\mu_1\braket{c_1}{c_2}},~~\frac{\gamma_1^2}{\gamma_2^2} =  \frac{\beta_2^2}{\beta_1^2} \frac{\mu_1^2}{\mu_2^2}.
\ee

So far we have analyzed the optimal distillation protocols under the constraint that only one branch leads to a GHZ. In what follows we will show that no distillation protocol can succeed with probability greater than that for OSBP, $P(\psi)$. In order to do so, we will study the behavior of $P(\psi)$ under LOCC, to conclude that it is a decreasing entanglement monotone. That is, given the state $\ket{\psi}$ and a sequence of local quantum operations that transform it into $\ket{\psi_i}$ with probability $p_i$, we will show that
\be
\label{monotone}
P(\psi)\geq\sum_ip_iP(\psi_i),
\ee
which means that the average probability to obtain a GHZ state from $\ket{\psi}$ using several branches is not greater than when using just one branch.

Although the set of transformations LOCC is very large and one should in principle check (\ref{monotone}) for any local protocol, we can use the fact that any such protocol decomposes into individual POVMs. Indeed, we need only prove the monotonic character of $P(\psi)$ under the most general local POVM on each subsystem. But, as a matter of fact, due to the symmetry of the problem it suffices to consider a general local POVM performed by one of the parties, say Alice. Notice, furthermore, that any POVM can be decomposed into a sequence of two-outcome POVMs \cite{POVM}.
Let us then consider a two-outcome POVM with operators $\{N_1,N_2\}$,  $N_1^{\dagger}N_1+N_2^{\dagger}N_2=\eins_A$, applied by Alice. With some probability $p_i$ the resulting state will be $\ket{\psi_i}\equiv p_i^{-1/2} N_i\otimes\eins_B\otimes\eins_C \ket{\psi}$, and then the parties can apply the optimal OSBP, with $A_i\otimes B_i \otimes C_i$ being the corresponding successful branch (see figure 2(i)). We want to show that
\be
P(\psi) \geq p_1P(\psi_1) + p_2P(\psi_2).
\label{condition1}
\ee
 Notice that the global action on Alice's side in order to distill a GHZ state can be reproduced by means of a single four-outcome POVM, namely by $\{A_1N_1, \bar{A}_1N_1, A_2N_2, \bar{A}_2N_2 \}$, where $\bar{A}_i$ are the operators satisfying $A_i^{\dagger}A_i+\bar{A}_i^{\dagger}\bar{A}_i =\eins_A$. The second and fourth operators disentangle the state. They are irrelevant for distillation, and we can join them together into some other operator $R$  (figure 2(ii)). Both operators $A_1N_1$ and $A_2N_2$, being the last transformation Alice applies to its subsystem, have to leave the local states orthonormal and therefore can be written as
\be
\label{povm2}
A_iN_i = \alpha_{1,i}\ket{0}\bra{\tilde{a_1}}+\alpha_{2,i}\ket{1}\bra{\tilde{a_2}}.
\ee
Consequently, this three-outcome POVM can be replaced with a two-outcome POVM with operators $R$ and 
\be
Q \equiv \sqrt{\alpha_{1,1}^2 + \alpha_{1,2}^2}\ket{0}\bra{\tilde{a}_1}+\sqrt{\alpha_{2,1}^2 + \alpha_{2,2}^2}\ket{1}\bra{\tilde{a}_2},
\ee
followed ---if the outcome corresponds to operator $Q$--- by a {\em diagonal}, two-outcome POVM (figure 2(iii)), 
\be
\label{povm4}
D_i=\frac{\alpha_{1,i}}{\sqrt{\alpha_{1,1}^2 + \alpha_{1,2}^2}}\proj{0}+\frac{\alpha_{2,i}}{\sqrt{\alpha_{2,1}^2 + \alpha_{2,2}^2}}\proj{1}.
\ee
Let $\ket{\phi^2}$ in (\ref{cas2}) be the normalized vector after applying $Q$ to $\ket{\psi}$. Then we can translate condition (\ref{condition1}) into
\be
P(\phi^2) \geq q_1P(\phi^2_1)+q_2P(\phi^2_2),
\label{condition2}
\ee
where $q_i$ is the probability of the outcome related to $D_i$ starting from $\ket{\phi^2}$, and $\ket{\phi^2_i}$ is the corresponding final state.

In order to finally check that (\ref{condition2}) holds, we further notice that the action of any two-outcome POVM on a given state $\ket{\psi}$ can be reproduced by a conditional series of {\em balanced}, two-outcome POVMs, namely POVMs such that, for each of them, the probability of the two outcomes is exactly $1/2$ \cite{POVM2}. That is, we only need to analyze a diagonal POVM where the square of the diagonal elements are $x/(2\mu_1^2)$ and $(1-x)/(2\mu_2^2)$ for the first operator $D_1$, where $x\in [2\mu_1^2-1,2\mu_1^2]$, and their completion to 1 for the second. Then (\ref{condition2}) is an inequality for $x$ that exhaustive numerical calculations have shown to saturate only for $x=\mu_1^2$, that is, when Alice acts trivially on her subsystem with $D_i=1/\sqrt{2}\eins_A$. Thus we conclude that the optimal probability for OSBP cannot be improved by any distillation protocol.

 Finally, we have also considered the optimal local approximate transformation \cite{Vid00} of $\ket{\psi}$ into a GHZ state. That is, given the state $\ket{\psi}$ and an arbitrary local protocol which transforms it into the (possibly mixed) state $\rho_i$ with probability $p_i$, we have looked for the maximal averaged fidelity between the output states $\rho_i$ and a GHZ state, 
\be
F_{opt} \equiv \max_{LOCC} \sum_i p_i \bra{GHZ}\rho_i\ket{GHZ}.
\label{fidelitat}
\ee
As we argued in the bipartite setting \cite{Vid00}, we only need to consider final pure states $\{\psi_i,p_i\}$ due to the manifest linearity of (\ref{fidelitat}) in $\rho_i$ and the possibility of {\em purifying} any protocol that produces mixed states. Exhaustive numerical search shows then that the optimal approximate transformation consists in a deterministic transformation of $\ket{\psi}$ into $\ket{\psi'}$ by means of local unitary transformations, exactly as in the bipartite case.

We have addressed the optimal distillation of a GHZ state starting from a pure state of a three-qubit system. We have shown that, contrary to what happens in the bipartite case, tripartite distillation requires that each party performs at least one local POVM. This feature can be associated to the lack of Schmidt decomposition for tripartite pure states. An alternative decomposition, based on non-orthogonal product states, has proven very useful to indicate which modifications must be introduced locally in a pure state in order to distill it into a GHZ state, and to optimize such distillation. 

We acknowledge financial support by the Austrian SF, by the Spanish MEC (AP98 and AP99) and by the European Community (ESF; TMR network ERB-FMRX-CT96-0087; project EQUIP; HPMF-CT-1999-00200). This work was concluded during the 2000 session of the Benasque Center for Science, Spain.

\begin{figure}
 \epsfysize=2cm
\begin{center}
 \epsffile{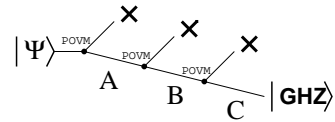}
\end{center}
 \caption{Scheme of a one succesful branch protocol (OSBP) for distillation of a GHZ state. \label{fig1}}
\end{figure}

\begin{figure}
 \epsfysize=5cm
\begin{center}
 \epsffile{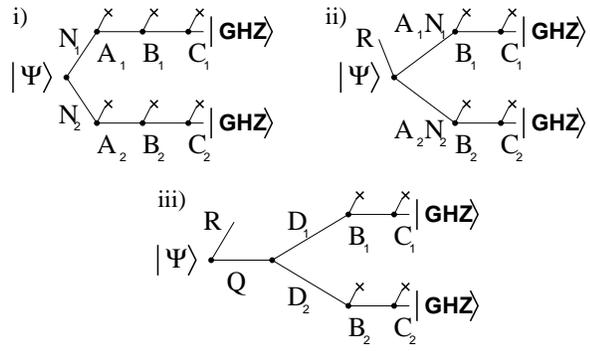}
\end{center}
 \caption{As far as distillation of a GHZ is concerned, these three series of conditional POVMs are equivalent (see equations (\ref{condition1}-\ref{condition2})) \label{fig2}}
\end{figure}


\begin{references}
\bibitem{Ben96} C.H. Bennett, H. Bernstein, S. Popescu and B. Schumacher, Phys. Rev. A{\bf 53} 3824 (1996).
\bibitem{Vid99} G. Vidal, Phys. Rev. Lett. {\bf 83} 1046 (1999). D. Jonathan and M.B. Plenio, Phys. Rev. Lett. {\bf 83}, 1455 (1999).
\bibitem{Dur00} W. D\"{u}r, G. Vidal and J.I. Cirac, quant-ph/0005115.
\bibitem{EPR} Einstein, Podolsky and Rosen, Phys. Rev. {\bf 47}, 777 (1935).
\bibitem{GHZ} D. M. Greenberger, M. Horne, A. Zeilinger, {\em Bell's theorem.}, ed. M. Kafatos, Kluwer, Dordrecht 69 (1989).
\bibitem{Gis98} N. Gisin and H. Beschmann-Pasquinucci, quant-ph/9804045.
\bibitem{Coh00} O. Cohen and T.A. Brun, Phys. Rev. Lett. {\bf 84}, 5908 (2000).
\bibitem{Vid00} G. Vidal, D. Jonathan and M. A. Nielsen, Phys. Rev. A 62, 012304 (2000).
\bibitem{des} A. Ac\'{\i}n, A. Andrianov, L. Costa, E. Jan\'e, J.I. Latorre and R. Tarrach, Phys. Rev. Lett. {\bf 85}, 1560 (2000). H. A. Carteret, A. Higuchi, A. Sudbery, quant-ph/0006125.
\bibitem{rab} See \cite{Ben96} for details on how this kind of POVMs can be implemented.
\bibitem{Lo97} H.-K. Lo and S. Popescu, quant-ph/9707038.
\bibitem{Vid99b} G. Vidal, Journ. of Mod. Opt. {\bf 47}, 355 (2000).  
\bibitem{phase} A third condition to be satisfied is that $\varphi+\varphi_a+\varphi_b+\varphi_c = 0 \mbox{ mod } 2\pi$, although this can be accomplished by a simple local unitary transformation once the local states $\ket{\kappa_i}$ have been orthonormalized, and it does not affect the probability of a successful distillation.
\bibitem{POVM} For instance, a three outcome POVM with operators $\{E_i\}_{i=1}^3$, $\sum_i E_i^{\dagger}E_i = \eins$, can be decomposed into a first two-outcome POVM with operators 
\be
\{~E_1, ~R\equiv \sqrt{E_2^{\dagger}E_2+E_3^{\dagger}E_3}~\}
\ee
and a second two-outcome POVM, to be applied only if the outcome of the first POVM corresponds to the operator $R$, with operators $\{E_2R^{-1}, E_3R^{-1}\}$.

\bibitem{POVM2} Consider an {\em unbalanced}, two-outcome POVM $\{E_1, E_2\}$, $E_1^{\dagger}E_1 + E_2^{\dagger}E_2=\eins$, applied to the state $\ket{\psi}$, with $q\geq1/2$ [$1-q$] being the probability of obtaining the outcome associated with $E_1$ $[E_2]$. Then the POVM
\be
\{~\frac{1}{\sqrt{2q}}E_1,~ R\equiv\sqrt{\frac{2q-1}{2q}E_1^{\dagger}E_1 + E_2^{\dagger}E_2}~\}
\label{crazy}
\ee
transforms the initial state according to $E_1$ with probability $1/2$. If the outcome corresponds to operator $R$, then a new 2-outcome, balanced POVM can be applied, which with probability $1/2$ will give one of the originally wished outcomes and with probability $1/2$ will require a next POVM. The whole series of balanced POVMs can be easily seen to converge towards the original unbalance POVM.
\end{references}
\end{document}